\documentclass[10pt,letterpaper]{article}

\usepackage{cogsci}

\cogscifinalcopy % Uncomment this line for the final submission 

\usepackage{pslatex}
\usepackage{apacite}
\usepackage{float}

\usepackage[flushleft]{threeparttable}
\usepackage{amsmath}
\usepackage{graphicx}
\usepackage{xcolor}

\title{A Rational Analysis of the Speech-to-Song Illusion}
 
\author{
{\large \bf Raja Marjieh$\textsuperscript{1}$,  Pol van Rijn$\textsuperscript{2}$, Ilia Sucholutsky$\textsuperscript{3}$, Harin Lee$\textsuperscript{2,4}$, Thomas L. Griffiths$\textsuperscript{1,3,*}$,  Nori Jacoby$\textsuperscript{2,*}$} \\
  $\textsuperscript{1}$Department of Psychology, Princeton University \\
  $\textsuperscript{2}$Computational Auditory Perception Group, Max Planck Institute for Empirical Aesthetics\\
  $\textsuperscript{3}$Department of Computer Science, Princeton University\\
  $\textsuperscript{4}$Max Planck Institute for Human Cognitive and Brain Sciences\\
  $\textsuperscript{*}$Equal Contribution\\
  {\small $\texttt{\{raja.marjieh, is2961, tomg\}@princeton.edu; \{pol.van-rijn, harin.lee, nori.jacoby\}@ae.mpg.de}$}  \\
  } 

\begin{document}

\maketitle

\begin{abstract}
The speech-to-song illusion is a robust psychological phenomenon whereby a spoken sentence sounds increasingly more musical as it is repeated. Despite decades of research, a complete formal account of this transformation is still lacking, and some of its nuanced characteristics, namely, that certain phrases appear to transform while others do not, is not well understood. Here we provide a formal account of this phenomenon, by recasting it as a statistical inference whereby a rational agent attempts to decide whether a sequence of utterances is more likely to have been produced in a song or speech. Using this approach and analyzing song and speech corpora, we further introduce a novel prose-to-lyrics illusion that is purely text-based. In this illusion, simply duplicating written sentences makes them appear more like song lyrics. We provide robust evidence for this new illusion in both human participants and large language models.

\end{abstract}

\section{Introduction}

First published as an audio demonstration just over two decades ago \cite{deutsch2003phantom, deutsch2019musical}, the speech-to-song illusion concerns a curious phenomenon whereby spoken phrases can be made to sound more song-like simply by repetition. This observation has been replicated and elaborated on in numerous studies \cite{deutsch2011illusory,tierney2013speech, tierney2018acoustic,tierney2018repetition, margulis2015pronunciation,simchy2018sound, falk2014speech,vanden2015everyday, rowland2019there}, and it taps into the deep relationship between speech and music which itself has been the center of considerable inquiry~\cite{zatorre2002structure, albouy2020distinct, ozaki2022globally, albouy2023spectro, ding2017temporal} tracing as far back as the works of eighteenth century philosophers such as Rousseu and Herder~\cite{rousseau1782collection,herder_2002}.

\begin{figure}[ht]
\centering
\includegraphics[width=0.9\linewidth]{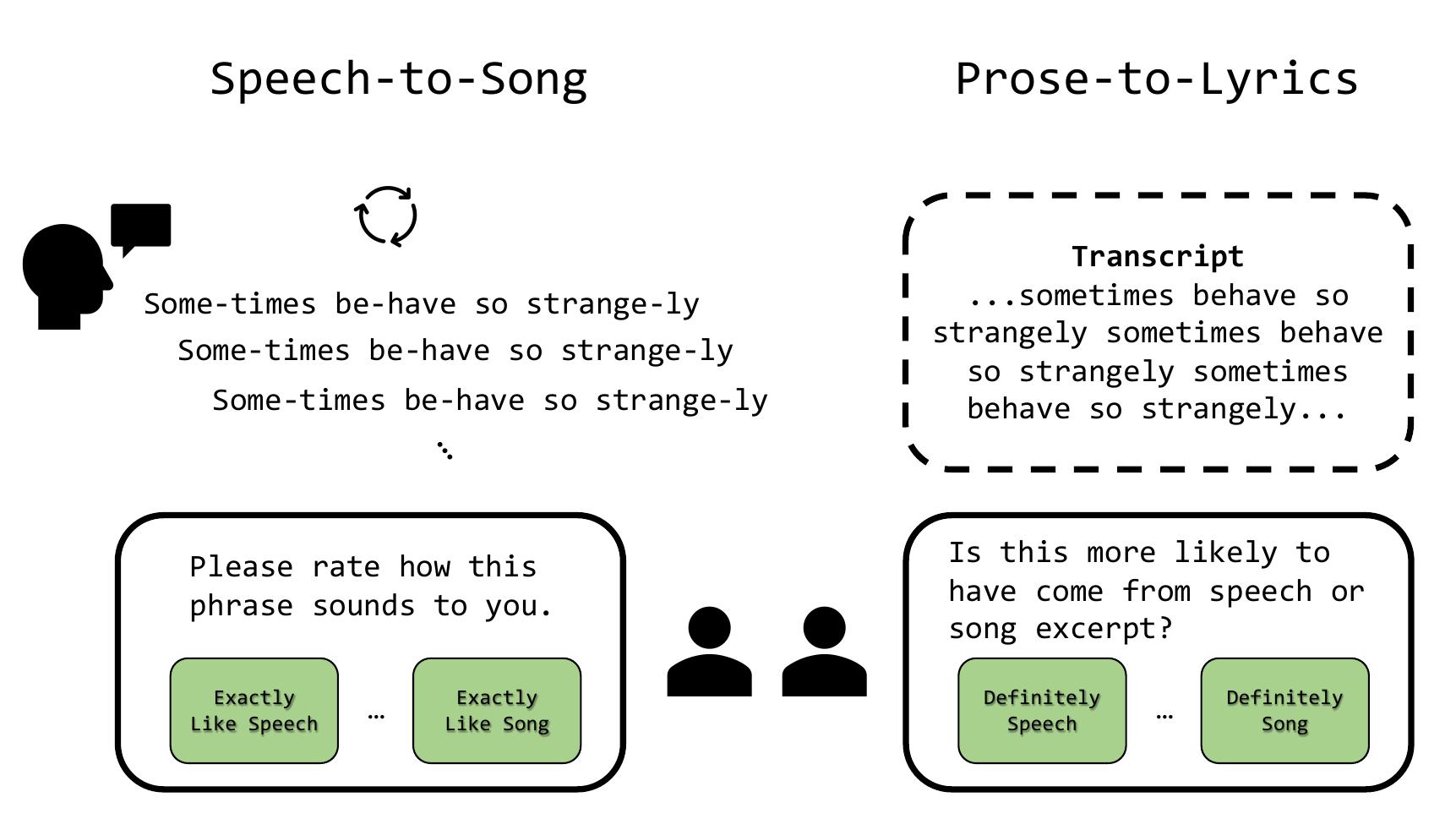}
\caption{The speech-to-song paradigm and the newly proposed prose-to-lyrics paradigm.}
\label{fig:schematic}
\end{figure}
Despite this rich research tradition, there is still no unifying theoretical framework to account for the full phenomenology of the illusion, namely, (i) what causes repetition on average to transform spoken phrases to song?~\cite{deutsch2011illusory} (ii) why does the strength of the effect increase with the number of repetitions?~\cite{tierney2018acoustic} (iii) why is it that certain phrases transform whereas others do not?~\cite{tierney2013speech} and (iv) how does this connect to the statistical properties of speech and music learned throughout life and culture? \cite{margulis2015pronunciation,jaisin2016speech} 

In the present work, we propose one such framework by analyzing the abstract computational problem underlying the speech-to-song illusion and recasting it as rational statistical inference \cite{anderson1990adaptive,griffiths2007mere}. Specifically, we argue that the phenomenon can be understood in terms of the extent to which a given stimulus (or its repetitions) is more likely to originate from a generative model of speech versus song. Using Bayesian inference, the problem of determining whether a stimulus is more speech-like or song-like reduces to analyzing the probability of that stimulus under suitable speech and song corpora. We show that textual transcription datasets of speech and songs, both naturalistic and synthetically generated by large language models (LLMs), are indeed sufficient to reproduce some of the key features of the illusion, and motivate a hypothesis regarding a new ``prose-to-lyrics'' illusion  that is based purely on text (i.e., no acoustic signal is provided) which we evaluate in both humans and LLMs (GPT-4; \citeNP{achiam2023gpt}). 

Our framework provides a principled account of the speech-to-song illusion that explicitly roots it in the learned statistics of speech and music, and allows for generalization across multiple modalities given suitable generative models. The paper proceeds as follows. We first review the empirical literature on the speech-to-song illusion, and the fundamentals of rational analysis in the context of a two-hypothesis decision problem. We then explicitly apply the latter to the speech-to-song problem and use it to motivate the textual version of the illusion. We then summarize the technical details of the corpus analysis and the human and LLM experiments, then present our results and distill their implications.

\section{Background}
%involves exposing participants to a spoken phrase (notably, "sometimes behave so strangely") which is repeated several times

\subsection{Speech-to-Song Illusion}
In its original form, the speech-to-song paradigm involves presenting participants with a spoken phrase (famously, ``sometimes behave so strangely'') and repeating it several times \cite{deutsch2011illusory, deutsch2019musical}. In the pause between each repetition, participants report how the last utterance sounds to them from  ``exactly like speech'' to ``exactly like song'' using a numeric Likert scale (Figure~\ref{fig:schematic}; Speech-to-Song). \citeA{deutsch2011illusory} showed that this simple manipulation is sufficient to drive participants to perceive the speech excerpt as more like song than speech relative to the original (and identical) presentation. In a subsequent experiment, the authors asked participants to repeat the utterance after one and ten presentations, and found that in the former case the acoustic characteristics of the participants' responses (their fundamental frequency $f_0$ contour) resembled those of speech utterances, whereas in the latter case those of a tonal melody. The fact that stimuli in all repetitions are identical is significant as it allows for studying the internal processing of speech and music without relying on different stimuli. 

Empirical research into the speech-to-song phenomenon has expanded substantially since the original study of~\citeA{deutsch2011illusory}. First, \citeA{margulis2015pronunciation} provided some evidence suggesting that the phenomenon may be affected by linguistic proficiency. Specifically, the authors showed that speech excerpts from languages that are hard to pronounce for native English speakers tended to transform more readily than those from languages that are easier to pronounce. Second, the increase of musicality of repeated sounds  has been shown to generalize to non-speech sounds such as random tone sequences~\cite{margulis2016repetition}, environmental sounds like dripping water~\cite{rowland2019there}, and animal sounds~\cite{simchy2018sound}. Third, while repetition tends to make stimuli sound more musical on average, it has been shown that the effect at the level of individual stimuli is more nuanced, with some transforming to song effectively while others resisting transformation altogether \cite{tierney2018acoustic}. Specifically, \citeA{tierney2013speech} meticulously scanned audiobooks in English to find ``song-like'' sentences that show enhanced effect of repetition. They then contrasted these stimuli with similar utterances that, without repetition, have comparable speech vs. song ratings, but do not exhibit significant transformation when repeated. 

Further research \cite{tierney2018acoustic,tierney2018repetition} revealed that this interaction also occurs when participants are presented with only the fundamental frequency ($f_0$) contour of a spoken utterance. However, simple acoustic manipulations of the contour (such as changing the beat consistency, pitch slope, and melodic structure) that were supposed to enhance the effect did not influence musicality ratings as expected. These findings highlight the complexity of extracting acoustic correlates to predict whether a given phrase should transform under repetition \cite{falk2014speech,tierney2018acoustic}.

\begin{figure*}[ht]
\centering
\includegraphics[width=0.65\linewidth]{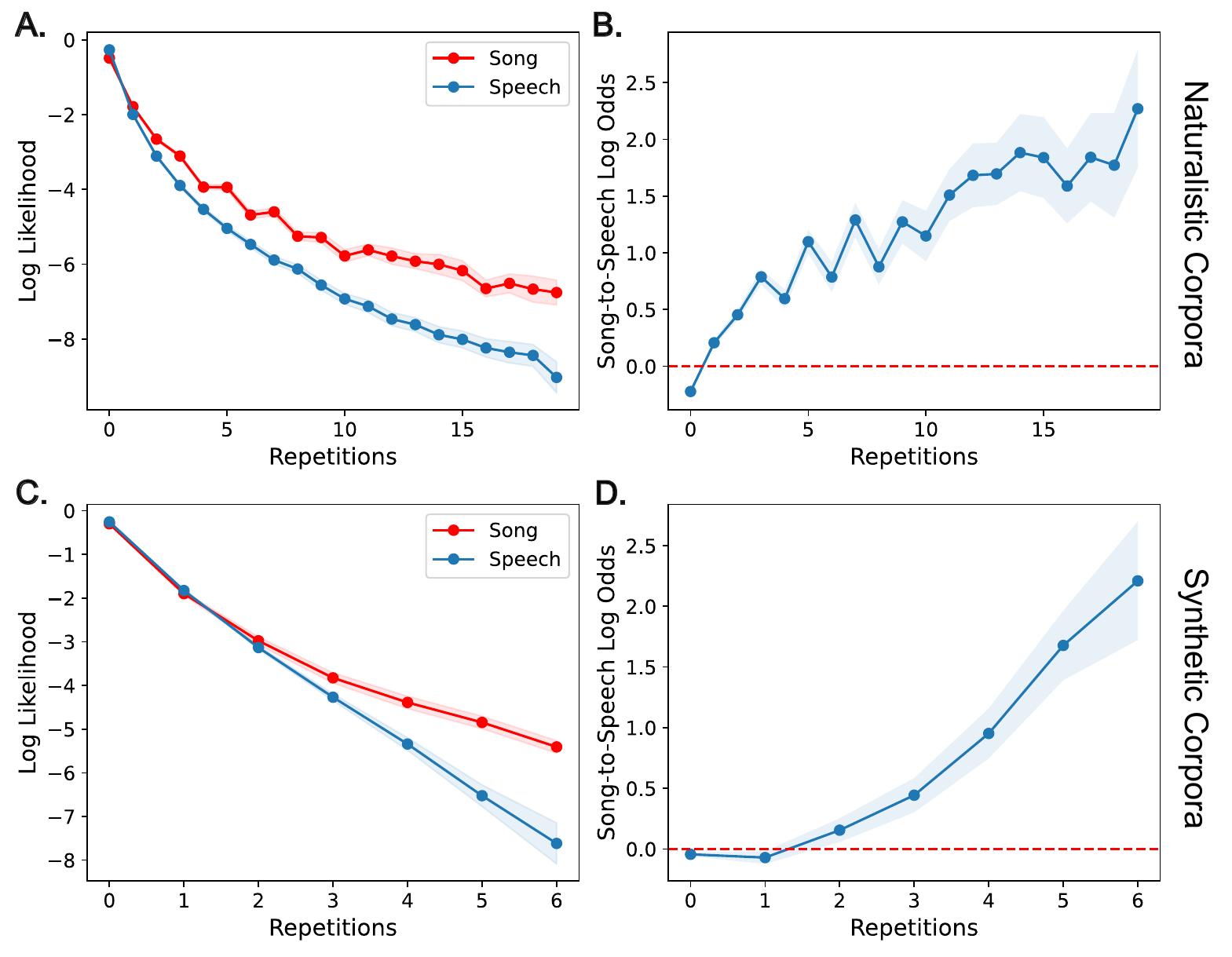}
\caption{Aggregate repetition analysis. \textbf{A.} Log-likelihood curves for the naturalistic dataset as a function of the number of repetitions (aggregated across all words). \textbf{B.} Log-odds ratio for the naturalistic dataset (positive values favor song). \textbf{C.} and \textbf{D.} Similar to \textbf{A} and \textbf{B} for the synthetic (LLM-generated) dataset.  Dashed red line indicates equal-likelihood threshold. Shaded area corresponds to 95\% confidence intervals bootstrapped over documents.}
\label{fig:log_odds_gpt_human}
\end{figure*}
\subsection{Rational Analysis}
Rational analysis is a cognitive modeling strategy that has been effectively applied across a wide array of topics including categorization \cite{anderson1990adaptive}, causal induction~\cite{griffiths2007mere}, perception~\cite{kersten2003bayesian}, and semantic memory~\cite{griffiths2007topics}. The core idea behind rational analysis is to analyze human behavior from the perspective of the abstract computational problem that it attempts to solve and the optimal solution to that problem~\cite{anderson1990adaptive}. This is usually implemented as Bayesian inference whereby the posterior probabilities of different hypotheses $p(h|d)$ are estimated by inverting a generative model of the data $d$ specified by some prior $p(h)$ and likelihood $p(d|h)$ using Bayes rule
\begin{equation}
    p(h|d)=\frac{p(d|h)p(h)}{\sum_{h}p(d|h)p(h)}
\end{equation}
The prior captures inductive bias towards certain hypotheses independent of the data, the likelihood captures how well hypotheses are supported by the data, and the posterior integrates the two. The main challenge in applying Bayes rule is computing the denominator as it often requires summing over a large number of hypotheses. In the context of binary decision problems, however, in which an agent needs to decide between two hypotheses $h_0$ and $h_1$, a simpler formula can be derived by taking the ratio between the posteriors of the two hypotheses. This is known as the log-odds form
\begin{equation}
    \log\frac{p(h_1|d)}{p(h_0|d)}=\log\frac{p(d|h_1)}{p(d|h_0)}+\log\frac{p(h_1)}{p(h_0)}
\end{equation}
From here we see that the extent to which a given hypothesis is preferred over the other ultimately depends on the likelihood ratio of the data as well as the prior odds.

\section{A Generative Account of Speech-to-Song}
We now leverage the two-hypothesis Bayesian log-odds formula to formulate a rational model of the speech-to-song task. The data in this case is the presented utterance $s$, or its $n$-repetitions, which we will denote by $s^n$. The log-odds formula for the speech-to-song task then becomes 
\begin{equation}\label{eqn:log-ratio}
    \log\frac{p(\text{song}|s^n)}{p(\text{speech}|s^n)}=\log\frac{p(s^n|\text{song})}{p(s^n|\text{speech})}+\log\frac{p(\text{song})}{p(\text{speech})}
\end{equation}
Note that the log prior odds is just a constant independent of $n$, and by design we can set it to zero (a priori speech and song stimuli are  equally likely). Thus, determining whether a sample is more speech-like or song-like becomes a matter of estimating the probability of the repeated stimulus $s^n$ under generative models of speech and song. We hypothesize that differences in the strength of transformation for different sentences will follow from the  statistics of those sentences under the different likelihood models. Equation \eqref{eqn:log-ratio} can also be used to predict the general effect of repetition by estimating $p(n|\text{speech})$ and $p(n|\text{song})$, i.e., the probability of observing an $n$-repetition in speech and song, respectively. 

Finding generative audio models to estimate the probability of $s^n$ is traditionally a hard problem (though recent advances in machine learning have changed this status; see Discussion). As an approximation, we can consider text corpora of speech and song transcripts. The idea here being that textual repetition in lyrics is indicative to some degree of repetition in melody (e.g., through refrains). By estimating the probability of sentences and their repetitions under such textual corpora we can evaluate whether their log-odds will exhibit the phenomenology of the illusion. Moreover, this treatment motivates the hypothesis that repetition should have a parallel effect in a purely textual variant of  the illusion (Figure \ref{fig:schematic}; Prose-to-Lyrics), as humans may also internalize similar statistical expectations in prose and lyrics. We will next present the modeling analysis and then proceed to the behavioral studies.

\section{Modeling Methods}
\begin{figure*}[ht]
\centering
\includegraphics[width=0.7\linewidth]{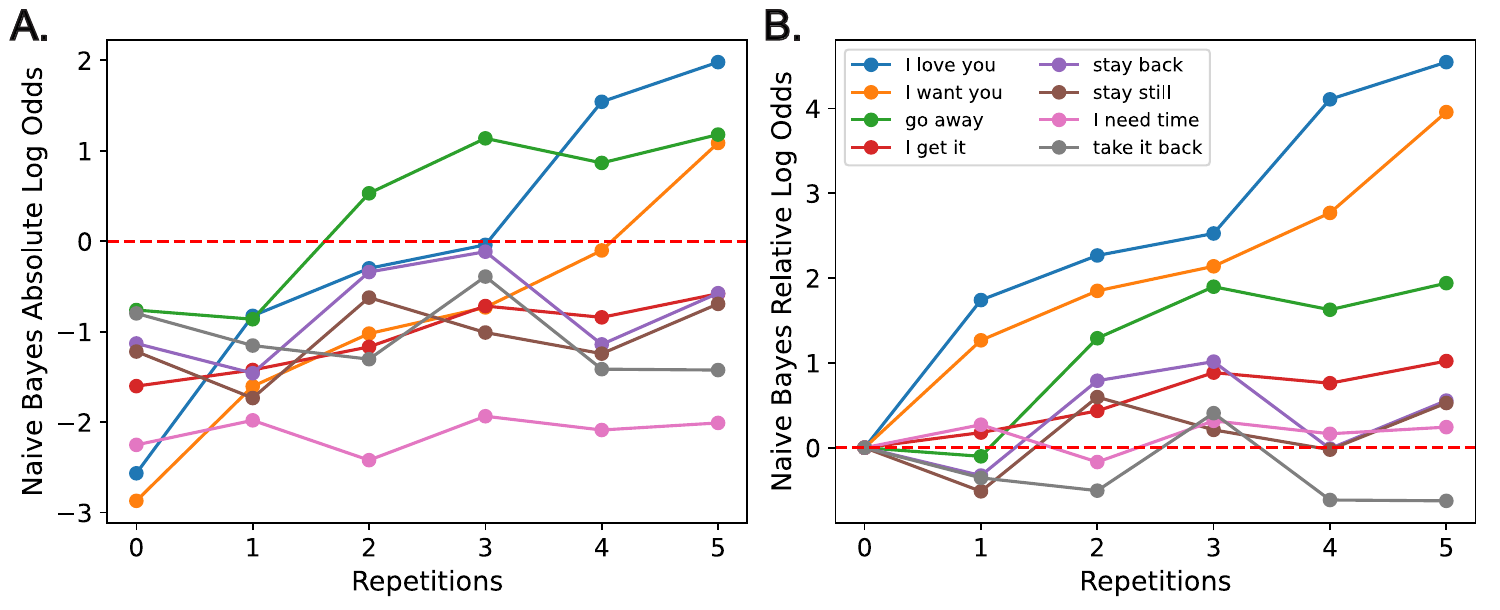}
\caption{Log-odds profiles for common sentences constructed from the naturalistic speech and song corpora under a na\"ive Bayes approximation. \textbf{A.} Absolute log-odds profiles, dashed red line indicates equal-likelihood threshold, positive values favor song. \textbf{B.} Relative log-odds profiles, with the value at zero subtracted (indicated by the dashed red line).}
\label{fig:sentence-dynamics}
\end{figure*}

\subsection{Text Preprocessing and Analysis}
To construct repetition probabilities from text corpora we applied a bag-of-words approach. Specifically, we measured the repetition of words within each text document irrespective of their location within the document. This is a simplifying assumption that increases statistical power and is common to other textual analysis methods (e.g. topic modeling; \citeNP{kherwa2019topic}). The processing steps were as follows. First, documents were tokenized, lemmatized, and cleaned from stop words and non-alphabetic entries using the \verb|nltk| Python package~\cite{bird2009natural}. Next, to create aggregate repetition probabilities per corpus, we counted the number of times each token was repeated within a given document, then created a histogram over the number of repetitions (i.e., the number of times a word was repeated $n$ times within a document), normalized, and finally averaged over all documents. A similar procedure was applied for word-level repetition probabilities (including pronouns), by counting the number of documents in which a given word was repeated $n$ times. Sentence-level repetition probabilities were constructed using a na\"ive Bayes approach (by multiplying the repetition probabilities of individual words). This is of course an approximation and can be generalized (see Discussion), but for the purpose of our theoretical analysis of short sentences constructed of common words it is sufficient. Specifically, the short sentences were constructed from the jointly most common words found in the naturalistic speech and song corpora (i.e., words that maximize the product of the number of documents from speech and song in which they appeared). These were useful as they allow for the construction of smooth word-level and sentence-level log-odds curves. Confidence intervals were constructed by bootstrapping over documents with replacement and 1,000 replicas.
\begin{figure*}[ht]
\centering
\includegraphics[width=0.65\linewidth]{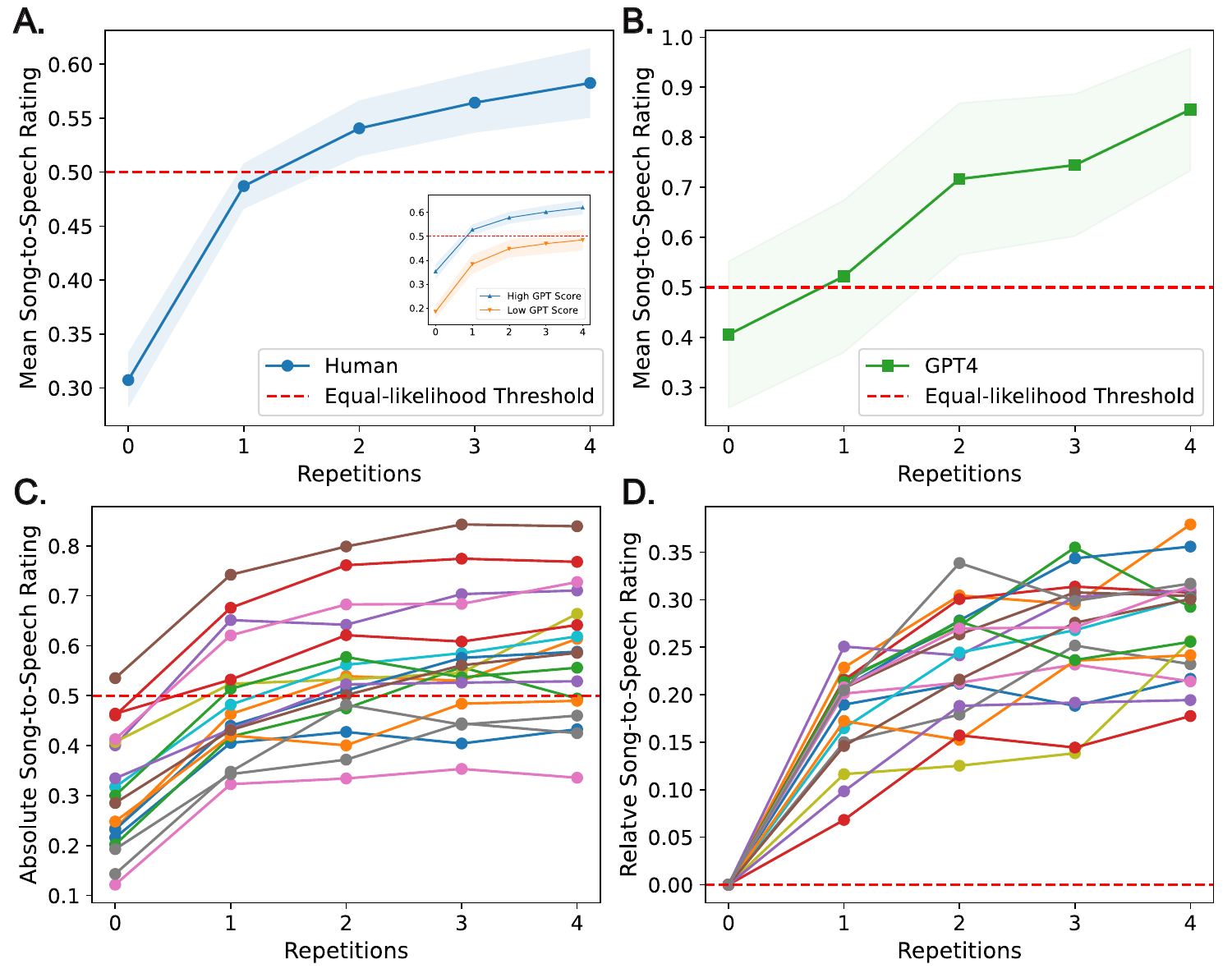}
\caption{Behavioral results of the prose-to-lyrics experiments. \textbf{A.} Mean human ratings as a function of repetitions (rescaled to a 0-1 range via division by a factor of 6; dashed red line indicates speech-song equality threshold). Inset shows the data grouped based on a median split of the GPT4 scores at 4 repetitions. Shaded area indicates 95\% confidence intervals (CIs) bootstrapped over participants. \textbf{B.} Mean rating as a function of repetition for the GPT4 control experiment. \textbf{C.} Mean sentence-level human ratings. \textbf{D.} Mean sentence-level human ratings relative to the value at 0 repetitions.}
\label{fig:prose-to-lyrics}
\end{figure*}

\subsection{Text Corpora for Speech and Song}
For our prose and lyrics corpora we relied on two complementary approaches: one where we mine text from an ecologically valid  (``naturalistic'') source, but accept the fact that different documents will have varied lengths, and another where we synthesize text artificially using a large language model (GPT4) but have full control over the document length. For the naturalistic song corpus, we used a collection of 2,994 song lyrics extracted from songs on Shazam, a mobile app for song identification. The subset was extracted from a larger collection at random, whereby the lyrical language was detected using the FastText algorithm~\cite{joulin2016fasttext}. For the speech corpus, we used text documents from the Corpus of Contemporary American English (COCA; \citeNP{davies2010corpus}), a large dataset constructed from a variety of sources such as magazines, blogs, and academic articles. Specifically, we focused on a set of 948 randomly sampled documents from the magazine subset of COCA. Note that while there were more song documents, these were naturally shorter, with a given document comprising 98 words on average as opposed to an average of 437 words for speech. To control for this variability, we used GPT4 to generate a synthetic set of 1,400 songs and 1,400 conversation transcripts, both constrained to approximately 20 lines per document, by repeatedly querying the model using the following prompt (with a temperature of 0.7 to encourage diversity in output): ``Please write 10 [song lyrics\textbar conversations] of length 20 lines each. Before each [song\textbar conversation], write [Song\textbar Conversation] and the number of the [song\textbar conversation]''.  The average number of tokens was indeed more aligned in this case and amounted to 54 (95\% CI: $[33,76]$) for song and 49 (95\% CI: $[32,66]$) for speech.

\section{Modeling Results}
We begin with an empirical evaluation of the properties of the Bayes log-odds formula as applied to the textual speech and song corpora. Our goal here is to see if, despite its simplicity and pure reliance on text, the model is able to capture the qualitative phenomenology of the speech-to-song illusion. Figure~\ref{fig:log_odds_gpt_human} shows the log-likelihoods and log-odds for both speech and song as a function of general word repetition (naturalistic: Figure~\ref{fig:log_odds_gpt_human}A-B, synthetic: Figure~\ref{fig:log_odds_gpt_human}C-D). We see that for both cases, the log-likelihoods of repetition decrease with the number of repetitions. Crucially, while these initially overlap for song and speech, the tails of the former are much heavier, driving the log odds formula to favor song within two repetitions. We also see that this difference in tails leads to a monotonic increase in the log-odds formula towards favoring song as a function of repetition. These results are consistent with the expectations for the speech-to-song phenomenology, namely, that repetition on average renders utterances more song-like~\cite{deutsch2011illusory}, and increasingly so as a function of repetition~\cite{tierney2013speech,deutsch2011illusory,margulis2016repetition}. Note that a richer model could also capture ceiling effects by mapping the unbounded log-odds to a finite interval (e.g., with a sigmoid).

Next, we computed repetition log-odds for short sentences constructed from common words in the naturalistic corpus for which smooth profiles can be reliably estimated (see Modeling Methods). The resulting profiles are shown in Figure~\ref{fig:sentence-dynamics}A. As is the case with Figure~\ref{fig:log_odds_gpt_human}, we see a general trend of transformation from speech to song under repetition. Indeed, plotting the profiles relative to their value without repetition (i.e., collapsing the origin to zero; Figure~\ref{fig:sentence-dynamics}B) reveals that seven out of eight sentences increased over their original value, and that the rate of change is different in different sentences, similar to findings from the literature~\cite{tierney2013speech,tierney2018acoustic,tierney2018repetition}. For example, we see that intuitive sentences like ``I love you'' and ``I want you'' transform at a faster rate than more generic ones like ``I get it'' and ``go away''. Overall, this provides a principled explanation for the observed variation in phrase transformation rate based on natural statistics. 

Having evaluated the qualitative predictions of the textual model, we are now ready to test whether these are realized quantitatively in the hypothesized prose-to-lyrics paradigm.

\section{Behavioral Methods}
\subsection{Behavioral Paradigm}
\textbf{Humans}\, For the prose-to-lyrics experiment, we recruited 120 UK participants from the online recruitment platform Prolific (\url{https://www.prolific.com/}). All participants provided informed consent prior to participation in accordance with the Max Planck Ethics Council (\#2021\_42). Upon providing informed consent, participants received the following instructions: `In this study you will be presented with text transcripts of different audio recordings and your task is to decide based on those transcripts whether the original audio is more likely to have been a song or a speech excerpt. For each transcript you will have 7 response options ranging from 0 (definitely speech) to 6 (definitely song).' Participants then rated up to 20 randomly chosen stimuli and responded to the prompt: `The following text transcript was extracted from an audio recording: $\langle \text{transcript} \rangle$. Based on this transcript alone, is the audio recording more likely to have been a speech or a song excerpt?' (Figure \ref{fig:schematic}; Prose-to-Lyrics). Overall, there were 18 sentences (see Stimuli) each repeated 0 to 4 times (overall 90 items), and each rating item received an average of 63 ratings (range: 61 -- 66). All items were shuffled and participants were randomly assigned items from this set.

\noindent\textbf{GPT4}\, The LLM control experiment was constructed in a similar fashion. We used Azure's OpenAI API to query GPT4 (with temperature set to zero; version 0613) with the following prompt: ``In what follows you will see a transcript of an audio recording and your task is to decide whether this transcript came from a song or a conversation. Please provide a numerical score between 0 and 1 where 0 is definitely conversation and 1 is definitely song. Do not provide any explanation or any other text. Transcript: $\langle\text{transcript}\rangle$. Answer:''.
\subsection{Stimuli}
We used a set of 18 sentences based on the following sources. First, short sentences constructed from common words in the naturalistic corpora (see Modeling Methods).
This set included, `I love you', `I get it', `I want you', `stay back', `stay still', `go away', `take it back', and `I need time'. Second, medium-length sentences that were randomly extracted from the speech corpus. These were `how fast does a Zamboni go?', `but once in a while you step on it from one end to the other', `if it takes an extra minute to do it, that's fine', `and wait 10 seconds or so between each bounce', `I drilled a hole through the top of my rack', `what kind of car would it be?', `I grew up watching them', and `it has really kept us growing'. Third, two additional sentences: `sometimes behave so strangely' for its historical significance in the original paradigm, and an additional technical sentence `I will head to the department store tomorrow'. Repeated versions were constructed by simply concatenating the sentences.
\newpage

\section{Behavioral Results}
Figure~\ref{fig:prose-to-lyrics} shows the behavioral results for the set of 18 sentences where each sentence was presented with 0-4 repetitions (see Behavioral Methods). Overall, the human average ratings over all sentences (Figure~\ref{fig:prose-to-lyrics}A) exhibited an excellent inter-rater reliability of $r=.96$ (95\% CI: $[.94,.98]$) computed via a split-half method and bootstrapped over participants. Inspecting Figure~\ref{fig:prose-to-lyrics}A, we see that repetition steadily transformed the sentences from initially being significantly speech-like with mean rating of $.31$ (95\% CI: $[.28,.33]$; with 0 being definitely speech and 1 being definitely song) to significantly being song-like with mean rating $.58$ (95\% CI: $[.55, .61]$). A similar pattern was observed in the case of GPT4 ratings which performed an equivalent task (Figure~\ref{fig:prose-to-lyrics}B). We also found that the human and LLM data were significantly correlated with a Pearson correlation of $r=.64$ (95\% CI: $[.60,.68]$). Moreover, splitting the human data into two groups based on a median split of the GPT4 scores at four repetitions (Figure~\ref{fig:prose-to-lyrics}A, inset) nicely separated the original curve into one that surpasses the 0.5 threshold and one that does not. These results show that we can replicate the effect of repetition in the speech-to-song illusion with our prose-to-lyrics illusion without directly playing an audio recording. 

\begin{figure}[ht]
\centering
\includegraphics[width=0.8\linewidth]{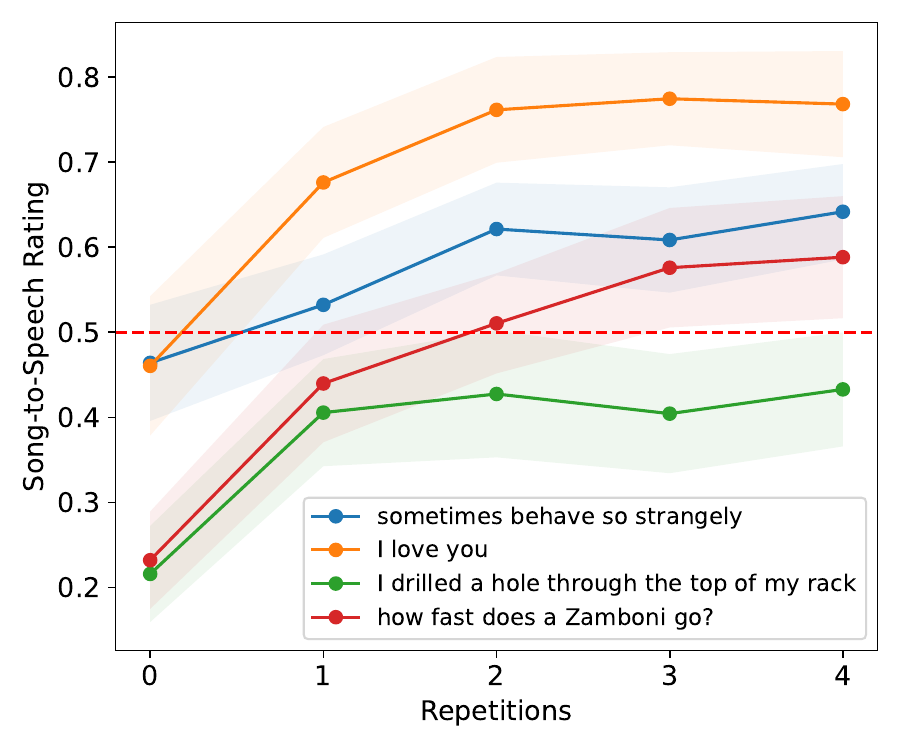}
\caption{Prose-to-lyrics illusion for demonstrative sentences. Lines represent averages, shaded area represents 95\% CIs.}
\label{fig:prose-to-lyrics-individual}
\end{figure}

Next, inspecting the sentence-level human mean ratings (Figure~\ref{fig:prose-to-lyrics}C), we see that the aggregate trend of monotonic increase in song ratings as a function of repetitions is also reflected at the level of individual sentences. This becomes particularly clear when the value at zero repetitions is subtracted (Figure~\ref{fig:prose-to-lyrics}D). We see that not only song ratings increase for all sentences, but they also do so at different rates as found for the speech-to-song illusion \cite{tierney2013speech,tierney2018repetition,tierney2018acoustic}. To further highlight the analogy to the speech-to-song illusion findings \cite{tierney2013speech}, we focus on a demonstrative subset of four sentences in Figure~\ref{fig:prose-to-lyrics-individual}. We see that while the pair `I love you' and `sometimes behave so strangely' are not significantly separated at zero repetitions ($.46$, CI: $[.38,.54]$, and $.46$, CI: $[.40,.53]$, respectively), the ratings of the former increase at a significantly faster rate (at 4 repetitions: $.77$, CI: $[.71,.83]$, and $.64$, CI: $[.59,.70]$, respectively). 
Likewise, while the pair `I drilled a hole through the top of my rack' and `how fast does a Zamboni go?' were significantly speech-like at zero repetitions and not significantly separated ($.22$, CI: $[.16,.27]$, and  $.23$, CI: $[.17,.29]$, respectively), only the latter transformed significantly to being song-like ($.43$, CI: $[.37,.50]$, and  $.59$, CI: $[.52,.66]$, respectively). We confirmed this sentence-repetition interaction using a non-parametric test ($p<0.001$ for both pairs). This provides clear evidence analogous to that found in the speech-to-song illusion \cite{tierney2013speech,tierney2018acoustic,tierney2018repetition}.

\section{Discussion}
By recasting the problem of deciding whether a stimulus is more like speech or song as rational statistical inference, we constructed a theoretical account of the speech-to-song illusion that not only provides principled explanations of its phenomenology based on the statistics of corpora, but also successfully predicts a new illusion that is purely textual. 

Our framework opens up multiple avenues for future investigation. First, a complete empirical treatment of the problem must ultimately use richer generative models of text and audio to estimate the probabilities of naturalistic stimuli and derive quantitative predictions. For text, we could train or refine LLMs on lyrics and prose, and then use that to estimate the probability of an entire (repeated) text conditioned on each subset. This should be possible with an open access language model like Llama 2~\cite{touvron2023llama}. For audio, we believe that we can also provide direct probability estimates of (repeated) utterances using modern audio transformer architectures such as the Music Transformer~\cite{huang2018music}. This task can be made easier by referring back to the observation of \citeA{tierney2018acoustic} that the speech-to-song effect can be replicated by providing only the $f_0$ contours of a sentence. This is significant because training a model only on $f_0$ contours of speech and sung utterances would require significantly less data than a full audio model. Second, finding balanced transforming and non-transforming illusion pairs is a hard problem~\cite{tierney2013speech}. Our theoretical account could potentially allow for the development of automated search algorithms for such pairs given suitable generative models. Third, being able to switch between generative models that are trained on music and speech corpora from different cultures could help develop principled hypotheses about the cross-cultural variation of this phenomenon~\cite{margulis2015pronunciation, jaisin2016speech}. For example, would it be possible to find phrases that transform in one language but not in another? We hope to engage with these ideas in future work.

The speech-to-song illusion will continue to fascinate listeners for decades to come. We believe that formalizing it as rational statistical inference whereby a perceived object provides sufficient evidence for an alternative, unconventional hypothesis serves as an exciting new step towards a computational theory of perceptual illusions.
\newpage
\section{Acknowledgments}
This work was supported by Microsoft Azure credits supplied to Princeton and by a Microsoft Foundation Models grant to TLG. The authors declare no competing interests.

\bibliographystyle{apacite}

\setlength{\bibleftmargin}{.125in}
\setlength{\bibindent}{-\bibleftmargin}

\bibliography{main}

\begin{thebibliography}{}

\bibitem [\protect \citeauthoryear {%
Achiam%
\ \protect \BOthers {.}}{%
Achiam%
\ \protect \BOthers {.}}{%
{\protect \APACyear {2023}}%
}]{%
achiam2023gpt}
\APACinsertmetastar {%
achiam2023gpt}%
\begin{APACrefauthors}%
Achiam, J.%
, Adler, S.%
, Agarwal, S.%
, Ahmad, L.%
, Akkaya, I.%
, Aleman, F\BPBI L.%
\BDBL {}others%
\end{APACrefauthors}%
\unskip\
\newblock
\APACrefYearMonthDay{2023}{}{}.
\newblock
{\BBOQ}\APACrefatitle {{GPT}-4 technical report} {{GPT}-4 technical report}.{\BBCQ}
\newblock
\APACjournalVolNumPages{arXiv preprint arXiv:2303.08774}{}{}{}.
\PrintBackRefs{\CurrentBib}

\bibitem [\protect \citeauthoryear {%
Albouy%
, Benjamin%
, Morillon%
\BCBL {}\ \BBA {} Zatorre%
}{%
Albouy%
\ \protect \BOthers {.}}{%
{\protect \APACyear {2020}}%
}]{%
albouy2020distinct}
\APACinsertmetastar {%
albouy2020distinct}%
\begin{APACrefauthors}%
Albouy, P.%
, Benjamin, L.%
, Morillon, B.%
\BCBL {}\ \BBA {} Zatorre, R\BPBI J.%
\end{APACrefauthors}%
\unskip\
\newblock
\APACrefYearMonthDay{2020}{}{}.
\newblock
{\BBOQ}\APACrefatitle {Distinct sensitivity to spectrotemporal modulation supports brain asymmetry for speech and melody} {Distinct sensitivity to spectrotemporal modulation supports brain asymmetry for speech and melody}.{\BBCQ}
\newblock
\APACjournalVolNumPages{Science}{367}{6481}{1043--1047}.
\PrintBackRefs{\CurrentBib}

\bibitem [\protect \citeauthoryear {%
Albouy%
, Mehr%
, Hoyer%
, Ginzburg%
\BCBL {}\ \BBA {} Zatorre%
}{%
Albouy%
\ \protect \BOthers {.}}{%
{\protect \APACyear {2023}}%
}]{%
albouy2023spectro}
\APACinsertmetastar {%
albouy2023spectro}%
\begin{APACrefauthors}%
Albouy, P.%
, Mehr, S\BPBI A.%
, Hoyer, R\BPBI S.%
, Ginzburg, J.%
\BCBL {}\ \BBA {} Zatorre, R\BPBI J.%
\end{APACrefauthors}%
\unskip\
\newblock
\APACrefYearMonthDay{2023}{}{}.
\newblock
{\BBOQ}\APACrefatitle {Spectro-temporal acoustical markers differentiate speech from song across cultures} {Spectro-temporal acoustical markers differentiate speech from song across cultures}.{\BBCQ}
\newblock
\APACjournalVolNumPages{bioRxiv}{}{}{2023--01}.
\PrintBackRefs{\CurrentBib}

\bibitem [\protect \citeauthoryear {%
Anderson%
}{%
Anderson%
}{%
{\protect \APACyear {1990}}%
}]{%
anderson1990adaptive}
\APACinsertmetastar {%
anderson1990adaptive}%
\begin{APACrefauthors}%
Anderson, J\BPBI R.%
\end{APACrefauthors}%
\unskip\
\newblock
\APACrefYear{1990}.
\newblock
\APACrefbtitle {The adaptive character of thought} {The adaptive character of thought}.
\newblock
\APACaddressPublisher{}{Psychology Press}.
\PrintBackRefs{\CurrentBib}

\bibitem [\protect \citeauthoryear {%
Bird%
, Klein%
\BCBL {}\ \BBA {} Loper%
}{%
Bird%
\ \protect \BOthers {.}}{%
{\protect \APACyear {2009}}%
}]{%
bird2009natural}
\APACinsertmetastar {%
bird2009natural}%
\begin{APACrefauthors}%
Bird, S.%
, Klein, E.%
\BCBL {}\ \BBA {} Loper, E.%
\end{APACrefauthors}%
\unskip\
\newblock
\APACrefYear{2009}.
\newblock
\APACrefbtitle {Natural language processing with Python: analyzing text with the natural language toolkit} {Natural language processing with python: analyzing text with the natural language toolkit}.
\newblock
\APACaddressPublisher{}{O'Reilly}.
\PrintBackRefs{\CurrentBib}

\bibitem [\protect \citeauthoryear {%
Davies%
}{%
Davies%
}{%
{\protect \APACyear {2010}}%
}]{%
davies2010corpus}
\APACinsertmetastar {%
davies2010corpus}%
\begin{APACrefauthors}%
Davies, M.%
\end{APACrefauthors}%
\unskip\
\newblock
\APACrefYearMonthDay{2010}{}{}.
\newblock
{\BBOQ}\APACrefatitle {The Corpus of Contemporary American English as the first reliable monitor corpus of {E}nglish} {The corpus of contemporary american english as the first reliable monitor corpus of {E}nglish}.{\BBCQ}
\newblock
\APACjournalVolNumPages{Literary and linguistic computing}{25}{4}{447--464}.
\PrintBackRefs{\CurrentBib}

\bibitem [\protect \citeauthoryear {%
Deutsch%
}{%
Deutsch%
}{%
{\protect \APACyear {2003}}%
}]{%
deutsch2003phantom}
\APACinsertmetastar {%
deutsch2003phantom}%
\begin{APACrefauthors}%
Deutsch, D.%
\end{APACrefauthors}%
\unskip\
\newblock
\APACrefYear{2003}.
\newblock
\APACrefbtitle {Phantom words and other curiosities} {Phantom words and other curiosities}.
\newblock
\APACaddressPublisher{}{Philomel Records}.
\PrintBackRefs{\CurrentBib}

\bibitem [\protect \citeauthoryear {%
Deutsch%
}{%
Deutsch%
}{%
{\protect \APACyear {2019}}%
}]{%
deutsch2019musical}
\APACinsertmetastar {%
deutsch2019musical}%
\begin{APACrefauthors}%
Deutsch, D.%
\end{APACrefauthors}%
\unskip\
\newblock
\APACrefYear{2019}.
\newblock
\APACrefbtitle {Musical illusions and phantom words: How music and speech unlock mysteries of the brain} {Musical illusions and phantom words: How music and speech unlock mysteries of the brain}.
\newblock
\APACaddressPublisher{}{Oxford University Press}.
\PrintBackRefs{\CurrentBib}

\bibitem [\protect \citeauthoryear {%
Deutsch%
, Henthorn%
\BCBL {}\ \BBA {} Lapidis%
}{%
Deutsch%
\ \protect \BOthers {.}}{%
{\protect \APACyear {2011}}%
}]{%
deutsch2011illusory}
\APACinsertmetastar {%
deutsch2011illusory}%
\begin{APACrefauthors}%
Deutsch, D.%
, Henthorn, T.%
\BCBL {}\ \BBA {} Lapidis, R.%
\end{APACrefauthors}%
\unskip\
\newblock
\APACrefYearMonthDay{2011}{}{}.
\newblock
{\BBOQ}\APACrefatitle {Illusory transformation from speech to song} {Illusory transformation from speech to song}.{\BBCQ}
\newblock
\APACjournalVolNumPages{The Journal of the Acoustical Society of America}{129}{4}{2245--2252}.
\PrintBackRefs{\CurrentBib}

\bibitem [\protect \citeauthoryear {%
Ding%
\ \protect \BOthers {.}}{%
Ding%
\ \protect \BOthers {.}}{%
{\protect \APACyear {2017}}%
}]{%
ding2017temporal}
\APACinsertmetastar {%
ding2017temporal}%
\begin{APACrefauthors}%
Ding, N.%
, Patel, A\BPBI D.%
, Chen, L.%
, Butler, H.%
, Luo, C.%
\BCBL {}\ \BBA {} Poeppel, D.%
\end{APACrefauthors}%
\unskip\
\newblock
\APACrefYearMonthDay{2017}{}{}.
\newblock
{\BBOQ}\APACrefatitle {Temporal modulations in speech and music} {Temporal modulations in speech and music}.{\BBCQ}
\newblock
\APACjournalVolNumPages{Neuroscience \& Biobehavioral Reviews}{81}{}{181--187}.
\PrintBackRefs{\CurrentBib}

\bibitem [\protect \citeauthoryear {%
Falk%
, Rathcke%
\BCBL {}\ \BBA {} Dalla~Bella%
}{%
Falk%
\ \protect \BOthers {.}}{%
{\protect \APACyear {2014}}%
}]{%
falk2014speech}
\APACinsertmetastar {%
falk2014speech}%
\begin{APACrefauthors}%
Falk, S.%
, Rathcke, T.%
\BCBL {}\ \BBA {} Dalla~Bella, S.%
\end{APACrefauthors}%
\unskip\
\newblock
\APACrefYearMonthDay{2014}{}{}.
\newblock
{\BBOQ}\APACrefatitle {When speech sounds like music.} {When speech sounds like music.}{\BBCQ}
\newblock
\APACjournalVolNumPages{Journal of Experimental Psychology: Human Perception and Performance}{40}{4}{1491}.
\PrintBackRefs{\CurrentBib}

\bibitem [\protect \citeauthoryear {%
Griffiths%
, Steyvers%
\BCBL {}\ \BBA {} Tenenbaum%
}{%
Griffiths%
\ \protect \BOthers {.}}{%
{\protect \APACyear {2007}}%
}]{%
griffiths2007topics}
\APACinsertmetastar {%
griffiths2007topics}%
\begin{APACrefauthors}%
Griffiths, T\BPBI L.%
, Steyvers, M.%
\BCBL {}\ \BBA {} Tenenbaum, J\BPBI B.%
\end{APACrefauthors}%
\unskip\
\newblock
\APACrefYearMonthDay{2007}{}{}.
\newblock
{\BBOQ}\APACrefatitle {Topics in semantic representation.} {Topics in semantic representation.}{\BBCQ}
\newblock
\APACjournalVolNumPages{Psychological Review}{114}{2}{211}.
\PrintBackRefs{\CurrentBib}

\bibitem [\protect \citeauthoryear {%
Griffiths%
\ \BBA {} Tenenbaum%
}{%
Griffiths%
\ \BBA {} Tenenbaum%
}{%
{\protect \APACyear {2007}}%
}]{%
griffiths2007mere}
\APACinsertmetastar {%
griffiths2007mere}%
\begin{APACrefauthors}%
Griffiths, T\BPBI L.%
\BCBT {}\ \BBA {} Tenenbaum, J\BPBI B.%
\end{APACrefauthors}%
\unskip\
\newblock
\APACrefYearMonthDay{2007}{}{}.
\newblock
{\BBOQ}\APACrefatitle {From mere coincidences to meaningful discoveries} {From mere coincidences to meaningful discoveries}.{\BBCQ}
\newblock
\APACjournalVolNumPages{Cognition}{103}{2}{180--226}.
\PrintBackRefs{\CurrentBib}

\bibitem [\protect \citeauthoryear {%
Herder%
}{%
Herder%
}{%
{\protect \APACyear {2002}}%
}]{%
herder_2002}
\APACinsertmetastar {%
herder_2002}%
\begin{APACrefauthors}%
Herder, J\BPBI G\BPBI v.%
\end{APACrefauthors}%
\unskip\
\newblock
\APACrefYearMonthDay{2002}{}{}.
\newblock
{\BBOQ}\APACrefatitle {Treatise on the Origin of Language (1772)} {Treatise on the origin of language (1772)}.{\BBCQ}
\newblock
\BIn{} M\BPBI N.~Forster\ (\BED), \APACrefbtitle {Herder: Philosophical Writings} {Herder: Philosophical writings}\ (\BPG~65–164).
\newblock
\APACaddressPublisher{}{Cambridge University Press}.
\PrintBackRefs{\CurrentBib}

\bibitem [\protect \citeauthoryear {%
Huang%
\ \protect \BOthers {.}}{%
Huang%
\ \protect \BOthers {.}}{%
{\protect \APACyear {2018}}%
}]{%
huang2018music}
\APACinsertmetastar {%
huang2018music}%
\begin{APACrefauthors}%
Huang, C\BHBI Z\BPBI A.%
, Vaswani, A.%
, Uszkoreit, J.%
, Shazeer, N.%
, Hawthorne, C.%
, Dai, A\BPBI M.%
\BDBL {}Eck, D.%
\end{APACrefauthors}%
\unskip\
\newblock
\APACrefYearMonthDay{2018}{}{}.
\newblock
{\BBOQ}\APACrefatitle {Music Transformer: Generating Music with Long-Term Structure} {Music transformer: Generating music with long-term structure}.{\BBCQ}
\newblock
\APACjournalVolNumPages{arXiv preprint arXiv:1809.04281}{}{}{}.
\PrintBackRefs{\CurrentBib}

\bibitem [\protect \citeauthoryear {%
Jaisin%
, Suphanchaimat%
, Figueroa~Candia%
\BCBL {}\ \BBA {} Warren%
}{%
Jaisin%
\ \protect \BOthers {.}}{%
{\protect \APACyear {2016}}%
}]{%
jaisin2016speech}
\APACinsertmetastar {%
jaisin2016speech}%
\begin{APACrefauthors}%
Jaisin, K.%
, Suphanchaimat, R.%
, Figueroa~Candia, M\BPBI A.%
\BCBL {}\ \BBA {} Warren, J\BPBI D.%
\end{APACrefauthors}%
\unskip\
\newblock
\APACrefYearMonthDay{2016}{}{}.
\newblock
{\BBOQ}\APACrefatitle {The speech-to-song illusion is reduced in speakers of tonal (vs. non-tonal) languages} {The speech-to-song illusion is reduced in speakers of tonal (vs. non-tonal) languages}.{\BBCQ}
\newblock
\APACjournalVolNumPages{Frontiers in Psychology}{7}{}{662}.
\PrintBackRefs{\CurrentBib}

\bibitem [\protect \citeauthoryear {%
Joulin%
\ \protect \BOthers {.}}{%
Joulin%
\ \protect \BOthers {.}}{%
{\protect \APACyear {2016}}%
}]{%
joulin2016fasttext}
\APACinsertmetastar {%
joulin2016fasttext}%
\begin{APACrefauthors}%
Joulin, A.%
, Grave, E.%
, Bojanowski, P.%
, Douze, M.%
, J{\'e}gou, H.%
\BCBL {}\ \BBA {} Mikolov, T.%
\end{APACrefauthors}%
\unskip\
\newblock
\APACrefYearMonthDay{2016}{}{}.
\newblock
{\BBOQ}\APACrefatitle {Fasttext. zip: Compressing text classification models} {Fasttext. zip: Compressing text classification models}.{\BBCQ}
\newblock
\APACjournalVolNumPages{arXiv preprint arXiv:1612.03651}{}{}{}.
\PrintBackRefs{\CurrentBib}

\bibitem [\protect \citeauthoryear {%
Kersten%
\ \BBA {} Yuille%
}{%
Kersten%
\ \BBA {} Yuille%
}{%
{\protect \APACyear {2003}}%
}]{%
kersten2003bayesian}
\APACinsertmetastar {%
kersten2003bayesian}%
\begin{APACrefauthors}%
Kersten, D.%
\BCBT {}\ \BBA {} Yuille, A.%
\end{APACrefauthors}%
\unskip\
\newblock
\APACrefYearMonthDay{2003}{}{}.
\newblock
{\BBOQ}\APACrefatitle {Bayesian models of object perception} {Bayesian models of object perception}.{\BBCQ}
\newblock
\APACjournalVolNumPages{Current opinion in neurobiology}{13}{2}{150--158}.
\PrintBackRefs{\CurrentBib}

\bibitem [\protect \citeauthoryear {%
Kherwa%
\ \BBA {} Bansal%
}{%
Kherwa%
\ \BBA {} Bansal%
}{%
{\protect \APACyear {2019}}%
}]{%
kherwa2019topic}
\APACinsertmetastar {%
kherwa2019topic}%
\begin{APACrefauthors}%
Kherwa, P.%
\BCBT {}\ \BBA {} Bansal, P.%
\end{APACrefauthors}%
\unskip\
\newblock
\APACrefYearMonthDay{2019}{}{}.
\newblock
{\BBOQ}\APACrefatitle {Topic modeling: a comprehensive review} {Topic modeling: a comprehensive review}.{\BBCQ}
\newblock
\APACjournalVolNumPages{EAI Endorsed transactions on scalable information systems}{7}{24}{}.
\PrintBackRefs{\CurrentBib}

\bibitem [\protect \citeauthoryear {%
Margulis%
\ \BBA {} Simchy-Gross%
}{%
Margulis%
\ \BBA {} Simchy-Gross%
}{%
{\protect \APACyear {2016}}%
}]{%
margulis2016repetition}
\APACinsertmetastar {%
margulis2016repetition}%
\begin{APACrefauthors}%
Margulis, E\BPBI H.%
\BCBT {}\ \BBA {} Simchy-Gross, R.%
\end{APACrefauthors}%
\unskip\
\newblock
\APACrefYearMonthDay{2016}{}{}.
\newblock
{\BBOQ}\APACrefatitle {Repetition enhances the musicality of randomly generated tone sequences} {Repetition enhances the musicality of randomly generated tone sequences}.{\BBCQ}
\newblock
\APACjournalVolNumPages{Music Perception: An Interdisciplinary Journal}{33}{4}{509--514}.
\PrintBackRefs{\CurrentBib}

\bibitem [\protect \citeauthoryear {%
Margulis%
, Simchy-Gross%
\BCBL {}\ \BBA {} Black%
}{%
Margulis%
\ \protect \BOthers {.}}{%
{\protect \APACyear {2015}}%
}]{%
margulis2015pronunciation}
\APACinsertmetastar {%
margulis2015pronunciation}%
\begin{APACrefauthors}%
Margulis, E\BPBI H.%
, Simchy-Gross, R.%
\BCBL {}\ \BBA {} Black, J\BPBI L.%
\end{APACrefauthors}%
\unskip\
\newblock
\APACrefYearMonthDay{2015}{}{}.
\newblock
{\BBOQ}\APACrefatitle {Pronunciation difficulty, temporal regularity, and the speech-to-song illusion} {Pronunciation difficulty, temporal regularity, and the speech-to-song illusion}.{\BBCQ}
\newblock
\APACjournalVolNumPages{Frontiers in Psychology}{6}{}{48}.
\PrintBackRefs{\CurrentBib}

\bibitem [\protect \citeauthoryear {%
Ozaki%
\ \protect \BOthers {.}}{%
Ozaki%
\ \protect \BOthers {.}}{%
{\protect \APACyear {2022}}%
}]{%
ozaki2022globally}
\APACinsertmetastar {%
ozaki2022globally}%
\begin{APACrefauthors}%
Ozaki, Y.%
, Tierney, A.%
, Pfordresher, P.%
, Mcbride, J.%
, Benetos, E.%
, Proutskova, P.%
\BDBL {}others%
\end{APACrefauthors}%
\unskip\
\newblock
\APACrefYearMonthDay{2022}{}{}.
\newblock
{\BBOQ}\APACrefatitle {Globally, songs and instrumental melodies are slower, higher, and use more stable pitches than speech [Stage 2 Registered Report]} {Globally, songs and instrumental melodies are slower, higher, and use more stable pitches than speech [stage 2 registered report]}.{\BBCQ}
\newblock

\PrintBackRefs{\CurrentBib}

\bibitem [\protect \citeauthoryear {%
Rousseau%
}{%
Rousseau%
}{%
{\protect \APACyear {1782}}%
}]{%
rousseau1782collection}
\APACinsertmetastar {%
rousseau1782collection}%
\begin{APACrefauthors}%
Rousseau, J\BHBI J.%
\end{APACrefauthors}%
\unskip\
\newblock
\APACrefYear{1782}.
\newblock
\APACrefbtitle {Collection compl{\`e}te des oeuvres de JJ Rousseau, citoyen de Gen{\`e}ve} {Collection compl{\`e}te des oeuvres de jj rousseau, citoyen de gen{\`e}ve}\ (\BVOL~3).
\PrintBackRefs{\CurrentBib}

\bibitem [\protect \citeauthoryear {%
Rowland%
, Kasdan%
\BCBL {}\ \BBA {} Poeppel%
}{%
Rowland%
\ \protect \BOthers {.}}{%
{\protect \APACyear {2019}}%
}]{%
rowland2019there}
\APACinsertmetastar {%
rowland2019there}%
\begin{APACrefauthors}%
Rowland, J.%
, Kasdan, A.%
\BCBL {}\ \BBA {} Poeppel, D.%
\end{APACrefauthors}%
\unskip\
\newblock
\APACrefYearMonthDay{2019}{}{}.
\newblock
{\BBOQ}\APACrefatitle {There is music in repetition: Looped segments of speech and nonspeech induce the perception of music in a time-dependent manner} {There is music in repetition: Looped segments of speech and nonspeech induce the perception of music in a time-dependent manner}.{\BBCQ}
\newblock
\APACjournalVolNumPages{Psychonomic Bulletin \& Review}{26}{}{583--590}.
\PrintBackRefs{\CurrentBib}

\bibitem [\protect \citeauthoryear {%
Simchy-Gross%
\ \BBA {} Margulis%
}{%
Simchy-Gross%
\ \BBA {} Margulis%
}{%
{\protect \APACyear {2018}}%
}]{%
simchy2018sound}
\APACinsertmetastar {%
simchy2018sound}%
\begin{APACrefauthors}%
Simchy-Gross, R.%
\BCBT {}\ \BBA {} Margulis, E\BPBI H.%
\end{APACrefauthors}%
\unskip\
\newblock
\APACrefYearMonthDay{2018}{}{}.
\newblock
{\BBOQ}\APACrefatitle {The sound-to-music illusion: Repetition can musicalize nonspeech sounds} {The sound-to-music illusion: Repetition can musicalize nonspeech sounds}.{\BBCQ}
\newblock
\APACjournalVolNumPages{Music \& Science}{1}{}{2059204317731992}.
\PrintBackRefs{\CurrentBib}

\bibitem [\protect \citeauthoryear {%
Tierney%
, Dick%
, Deutsch%
\BCBL {}\ \BBA {} Sereno%
}{%
Tierney%
\ \protect \BOthers {.}}{%
{\protect \APACyear {2013}}%
}]{%
tierney2013speech}
\APACinsertmetastar {%
tierney2013speech}%
\begin{APACrefauthors}%
Tierney, A.%
, Dick, F.%
, Deutsch, D.%
\BCBL {}\ \BBA {} Sereno, M.%
\end{APACrefauthors}%
\unskip\
\newblock
\APACrefYearMonthDay{2013}{}{}.
\newblock
{\BBOQ}\APACrefatitle {Speech versus song: multiple pitch-sensitive areas revealed by a naturally occurring musical illusion} {Speech versus song: multiple pitch-sensitive areas revealed by a naturally occurring musical illusion}.{\BBCQ}
\newblock
\APACjournalVolNumPages{Cerebral Cortex}{23}{2}{249--254}.
\PrintBackRefs{\CurrentBib}

\bibitem [\protect \citeauthoryear {%
Tierney%
, Patel%
\BCBL {}\ \BBA {} Breen%
}{%
Tierney%
\ \protect \BOthers {.}}{%
{\protect \APACyear {2018}}%
{\protect \APACexlab {{\protect \BCnt {1}}}}}]{%
tierney2018acoustic}
\APACinsertmetastar {%
tierney2018acoustic}%
\begin{APACrefauthors}%
Tierney, A.%
, Patel, A\BPBI D.%
\BCBL {}\ \BBA {} Breen, M.%
\end{APACrefauthors}%
\unskip\
\newblock
\APACrefYearMonthDay{2018{\protect \BCnt {1}}}{}{}.
\newblock
{\BBOQ}\APACrefatitle {Acoustic foundations of the speech-to-song illusion.} {Acoustic foundations of the speech-to-song illusion.}{\BBCQ}
\newblock
\APACjournalVolNumPages{Journal of Experimental Psychology: General}{147}{6}{888}.
\PrintBackRefs{\CurrentBib}

\bibitem [\protect \citeauthoryear {%
Tierney%
, Patel%
\BCBL {}\ \BBA {} Breen%
}{%
Tierney%
\ \protect \BOthers {.}}{%
{\protect \APACyear {2018}}%
{\protect \APACexlab {{\protect \BCnt {2}}}}}]{%
tierney2018repetition}
\APACinsertmetastar {%
tierney2018repetition}%
\begin{APACrefauthors}%
Tierney, A.%
, Patel, A\BPBI D.%
\BCBL {}\ \BBA {} Breen, M.%
\end{APACrefauthors}%
\unskip\
\newblock
\APACrefYearMonthDay{2018{\protect \BCnt {2}}}{}{}.
\newblock
{\BBOQ}\APACrefatitle {Repetition enhances the musicality of speech and tone stimuli to similar degrees} {Repetition enhances the musicality of speech and tone stimuli to similar degrees}.{\BBCQ}
\newblock
\APACjournalVolNumPages{Music Perception: An Interdisciplinary Journal}{35}{5}{573--578}.
\PrintBackRefs{\CurrentBib}

\bibitem [\protect \citeauthoryear {%
Touvron%
\ \protect \BOthers {.}}{%
Touvron%
\ \protect \BOthers {.}}{%
{\protect \APACyear {2023}}%
}]{%
touvron2023llama}
\APACinsertmetastar {%
touvron2023llama}%
\begin{APACrefauthors}%
Touvron, H.%
, Martin, L.%
, Stone, K.%
, Albert, P.%
, Almahairi, A.%
, Babaei, Y.%
\BDBL {}others%
\end{APACrefauthors}%
\unskip\
\newblock
\APACrefYearMonthDay{2023}{}{}.
\newblock
{\BBOQ}\APACrefatitle {Llama 2: Open foundation and fine-tuned chat models} {Llama 2: Open foundation and fine-tuned chat models}.{\BBCQ}
\newblock
\APACjournalVolNumPages{arXiv preprint arXiv:2307.09288}{}{}{}.
\PrintBackRefs{\CurrentBib}

\bibitem [\protect \citeauthoryear {%
Vanden Bosch~der Nederlanden%
, Hannon%
\BCBL {}\ \BBA {} Snyder%
}{%
Vanden Bosch~der Nederlanden%
\ \protect \BOthers {.}}{%
{\protect \APACyear {2015}}%
}]{%
vanden2015everyday}
\APACinsertmetastar {%
vanden2015everyday}%
\begin{APACrefauthors}%
Vanden Bosch~der Nederlanden, C\BPBI M.%
, Hannon, E\BPBI E.%
\BCBL {}\ \BBA {} Snyder, J\BPBI S.%
\end{APACrefauthors}%
\unskip\
\newblock
\APACrefYearMonthDay{2015}{}{}.
\newblock
{\BBOQ}\APACrefatitle {Everyday musical experience is sufficient to perceive the speech-to-song illusion.} {Everyday musical experience is sufficient to perceive the speech-to-song illusion.}{\BBCQ}
\newblock
\APACjournalVolNumPages{Journal of experimental psychology: General}{144}{2}{e43}.
\PrintBackRefs{\CurrentBib}

\bibitem [\protect \citeauthoryear {%
Zatorre%
, Belin%
\BCBL {}\ \BBA {} Penhune%
}{%
Zatorre%
\ \protect \BOthers {.}}{%
{\protect \APACyear {2002}}%
}]{%
zatorre2002structure}
\APACinsertmetastar {%
zatorre2002structure}%
\begin{APACrefauthors}%
Zatorre, R\BPBI J.%
, Belin, P.%
\BCBL {}\ \BBA {} Penhune, V\BPBI B.%
\end{APACrefauthors}%
\unskip\
\newblock
\APACrefYearMonthDay{2002}{}{}.
\newblock
{\BBOQ}\APACrefatitle {Structure and function of auditory cortex: music and speech} {Structure and function of auditory cortex: music and speech}.{\BBCQ}
\newblock
\APACjournalVolNumPages{Trends in Cognitive Sciences}{6}{1}{37--46}.
\PrintBackRefs{\CurrentBib}

\end{thebibliography}

\end{document}